\documentclass[aps,twocolumn,pra,tightenlines,floatfix,showpacs,superscriptaddress]{revtex4-1}
\usepackage{graphicx}
\usepackage{amsmath}
\usepackage{amssymb}
\usepackage{times}
\usepackage{epstopdf}
\usepackage[english]{babel}
\usepackage{natbib}
\usepackage{hyperref}

\begin{document}
\title{Matter-wave propagation in optical lattices: geometrical and flat-band effects}
\author{Mekena Metcalf}
\affiliation{School of Natural Sciences, University of California, Merced, CA 95343, USA}
\author{Gia-Wei Chern}
\affiliation{Theoretical Division and Center for Nonlinear Science, Los Alamos National Laboratory, Los Alamos, NM 87545}
\author{Massimiliano Di Ventra}
\affiliation{Department of Physics, University of California, San Diego, La Jolla, CA 92093, USA}
\author{Chih-Chun Chien}
\affiliation{School of Natural Sciences, University of California, Merced, CA 95343, USA}
\email{cchien5@ucmerced.edu}

\begin{abstract}
The geometry of optical lattices can be engineered allowing the study of atomic transport along paths 
arranged in patterns that are otherwise difficult to probe in the solid state. A question feasible to atomic systems is related to the speed of propagation of matter-waves as a function of the lattice geometry. To address
this issue, we have investigated theoretically the quantum transport of non-interacting and weakly-interacting ultracold fermionic atoms in several 2D optical lattice geometries. We find that the triangular lattice has a higher propagation velocity compared to the square lattice, and the cross-linked square lattice has an even faster propagation velocity. The increase results from the mixing of the momentum states which leads to different group velocities in quantum systems. Standard band theory provides an explanation and allows for a systematic way to search and design systems with controllable matter-wave propagation. Moreover, the presence of a flat band such as in a two-leg ladder geometry leads to a dynamical density discontinuity due to its localized atoms. Possible realizations of those dynamical phenomena are discussed.
\end{abstract}


\maketitle

\section{Introduction}
Variational calculus leads to the conclusion that the shortest distance a free classical particle follows between two points is a geodesic \cite{LeonhardtBook}. Is this also true for quantum particles? As expected the answer is not trivial. For instance, in the studies of quantum dynamics of magnetic domains, it has been found that interesting patterns emerge \cite{Subedi13}, and an analogue of a quantum-mechanical ``forest fire propagation'' has been realized \cite{Park13}.
Moreover, when the classical random-walk problem is promoted to a quantum one, the dynamics is no longer diffusive and the object spreads faster (see \cite{QW_book} and references therein). Therefore, quantum effects, in particular the wave nature, can alter the dynamics in a fundamental way. The detection of such unusual behavior can be, however, difficult in the solid state. To measure matter waves representing electrons in metals, attosecond spectroscopy needs to be implemented and the observation length scale is limited by scattering effects \cite{Neppl15}. Recent demonstrations of controlling group velocity of cold-atom matter waves \cite{LaserBEC_GV} and photons \cite{PhotonMolecule} further illustrate the feasibility of tuning group velocity in quantum systems.

Ultracold atoms in engineered optical lattices allow direct measurement of matter-wave dynamics. Many interesting 2D lattice geometries, including square, triangle, honeycomb, kagome, have been fabricated \cite{Greiner02,Struck11,Panahi11,Tarruell12,Jo12}. These systems have been shown to be versatile quantum simulators of complex many-body systems \cite{Qsim1,Qsim3,Qsim2,Qsim4} because of the wide selections of atomic species and tunable parameters such as interactions or trapping potentials. Recent advances in introducing time dependence to the parameters and configurations of cold-atom systems open up opportunities for studying nonequilibrium physics, particularly transport phenomena \cite{CAdyn2,CAdyn3,CAdyn4,Chien14,ChienReview}. Importantly, while the Fermi velocity, $v_F$, of electrons in typical metals such as copper is on the order of $10^6$ m/s, in cold-atom systems $v_F$ is on the order of $10^{-3}$ m/s \cite{OpticalMetamat}. Such a slow motion of cold atoms then allows detailed analyses of their dynamics.

As illustrated by the matter-wave example, transport of fermionic atoms is closely related to electronic transport in nanoscale and mesoscopic systems \cite{Brantut12,Brantut13,Chien14,ChienReview}. For instance, when ultracold fermions are driven out of equilibrium in a one-dimensional optical lattice, a quasi-steady state current with a constant magnitude for a period of time emerges,  which is the precursor of the steady-state current (in the thermodynamic limit) found in biased solid-state systems  \cite{MCFshort,MCF_TD}. Quasi-steady-state currents have also been found to survive in two-dimensional systems \cite{Chern14}. Interestingly, this quasi-steady state persists also in the absence of particle interactions, a
fact not easily verifiable in the solid state, while the presence of strong interactions can change the transport from ballistic to diffusive \cite{Schneider12}. 

Simulating band-theory of electrons using cold-atoms has made it possible to unambiguously demonstrate paradigmatic phenomena. For example, Bloch oscillations are known in band theory but defied a clear demonstration in solid state systems until their demonstrations in cold-atom systems (see~\cite{Mahan_CondMatBook} and references therein). It is thus reasonable to expect that cold-atom systems can provide more insights into geometric effects on matter-wave propagation. Since a mass current of cold atoms corresponds to a traveling matter wave, it is interesting to clarify how fast the wavefront propagates, and how its speed can be controlled, thus providing an answer to the question posited at the beginning. 

To explore the dependence of group velocity on the underlying geometry, we have employed noninteracting and weakly-interacting fermions in various optical lattice geometries and found that quantum matter waves can be accelerated altering the underlying lattice geometry, which is not expected in classical particles. Using band theory in the thermodynamic limit, we found that interference of matter waves plays a key role in the atom dynamics. This also provides guidance for finding lattice geometries with faster matter-wave propagation. Including a weak repulsive interaction at the mean-field level does not change the conclusions qualitatively.

We also note that certain lattice geometries can support flat bands, which refer to a special class of dispersionless bands (see Ref.~\cite{LiuCPB14,Wu07,Wu08} and references therein). For example, the kagome lattice can support a flat band and can be realized in optical lattices \cite{Jo12}. The particles residing on a flat band do not possess kinetic energy and as a consequence, they do not participate directly to transport and lead to interesting phenomena. For instance, a dynamically generated flat-band insulator sustaining a density discontinuity has been predicted in optical kagome lattices \cite{Chern14}. We find transport features similar to this flat band effect in a two-leg ladder, and discuss its possible experimental realization. Such examples suggest that not only the mobile properties but also interesting insulating ones could be explored using cold atoms in engineered lattice potentials. 

In addition, the zig-zag lattice provides an opportunity to explore rich physics related to frustrations in optical lattice systems \cite{ZigZag,Jo15}. One important feature of the zig-zag lattice is that the roles of the nearest neighbors and the next nearest neighbors can be switched by tuning the coupling or lattice parameters. We explore the same idea in transport by considering noninteracting fermions in a zig-zag lattice with tunable tunneling coefficients and show that the speed of  matter-wave propagation can be controlled. A maximal velocity emerges as the relative tunneling coefficients are tuned continuously.

This paper is organized as follows. In Sec.~\ref{sec:theory} the lattice geometries and models for studying matter-wave propagation are introduced, and band theory is briefly reviewed. We present the geometry-dependent matter-wave velocity and flat-band induced density discontinuity, along with their explanations from band theory in the thermodynamic limit, in Sec.~\ref{sec:result}. Sec.~\ref{sec:exp} discusses possible experimental setups for verifying the results and applications in controlling matter-wave propagation. Then Sec.~\ref{sec:conclusion} concludes this work.

\section{Theoretical background}\label{sec:theory}
\begin{figure}
\includegraphics[width=3.4in,clip]{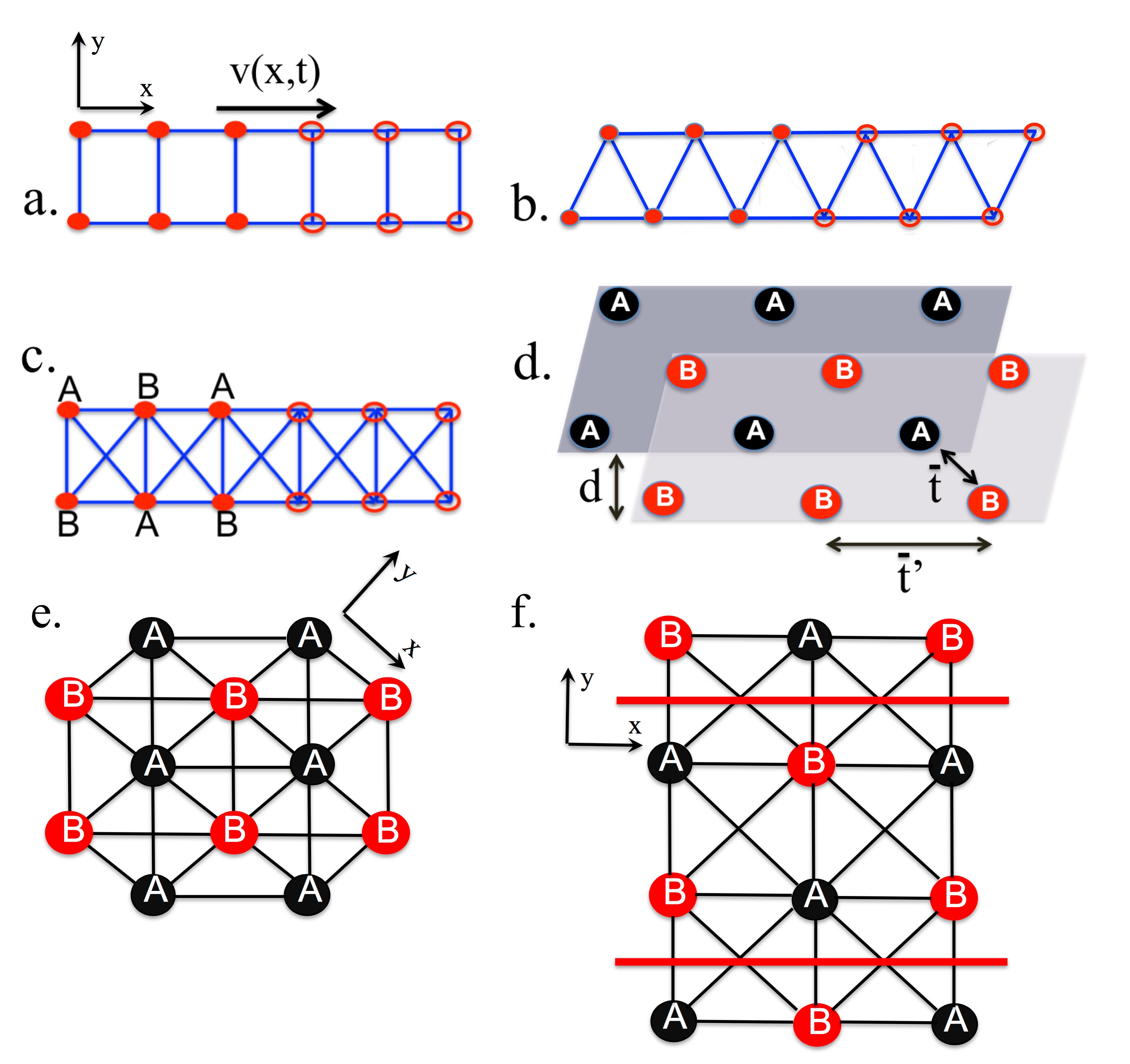}
\caption{(Color online) Illustration of the lattice geometries: (a) Two-leg square lattice. The coordinates $x,y$ are also shown. (b) Two-leg triangular lattice. (c) Two-leg square lattice with next-nearest neighbor hopping (SNNN) lattice with the sub-lattices A and B labeled. The hoppings between the A-A, A-B, and B-B links are all allowed in the SNNN lattice. Initially the left half of the lattice is filled (solid  circles) and the right half is empty (empty circles). The density imbalance drives a matter-wave propagating to the right-hand side, indicated by the velocity vector. (d) Possible experimental realization of the SNNN lattice with tunable tunnelings using bilayered optical lattices. One layer has a square lattice of the A-sites and the other layer has a square lattice of the B-sites. When viewed from the top of the bilayer lattice, inter- and intra- layer tunneling constructs the A-A, A-B, and B-B links as shown in (e). (f) By rotating the geometry of (e) by $\pi/4$ and using two laser sheets (denoted by the thick red lines) to confine a two-leg region, the SNNN lattice of (c) can be realized.}
\label{Fig:Illustration}
\end{figure}

To highlight interesting geometrical effects on quantum transport, we consider three types of 2D lattices illustrated in Figure~\ref{Fig:Illustration}. They are the square lattice, triangular lattice, and square lattice with next-nearest neighbor hopping (the SNNN lattice). The SNNN lattice can be viewed as a Bravais lattice with two lattice sites per unit cell, which we may call A and B. In the SNNN lattice both the nearest neighbor (A-A and B-B) and next-nearest neighbor (A-B) hoppings are allowed. One may consider the SNNN lattice as the 2D version of the body-centered cubic lattice in 3D \cite{AshcroftBook} because the SNNN lattice can be constructed by two intercalating square sub-lattices (labeled as A and B). The SNNN lattice is also geometrically similar to the checkerboard lattice demonstrated in Ref.~\cite{Tarruell12}. In our discussion the relative hopping coefficients of the A-A, A-B, and B-B links are assumed to be tunable. A possible experimental realization may use bilayered square lattices as shown in Fig.~\ref{Fig:Illustration}(d), although the need for a phase shift between the two layers could be a challenge. By viewing perpendicularly into the bilayer lattice, the two square sublattices (A and B) are intercalated as depicted by the SNNN lattice explained in Figure~\ref{Fig:Illustration} (e) and (f). The lattice constant $a$ serving as the length unit is chosen to be the same for the square and triangular lattice and the A-B link in the SNNN lattice.

For noninteracting single-component fermions in a moderate lattice potential, the system may be modeled by a tight-binding Hamiltonian of the form \cite{Hofstetter02}
\begin{equation}
H = - \sum_{\langle ij\rangle} \bar{t}_{ij} c^{\dagger}_i c_{j},
\end{equation}
where $\langle ij\rangle$ denotes a pair of sites connected by a link, $c_{i}$ ($c^{\dagger}_{i}$) annihilates (creates) a fermion at site $i$, and $\bar{t}_{ij}$ is the hopping coefficient. For a uniform lattice with $\bar{t}_{ij}=\bar{t}$, the unit of time is defined as $t_0 = \hbar / \bar{t}$. We set $\hbar = 1$ and assume that there are $L_x$ ($L_y$) lattice sites along the horizontal (vertical) direction. Recently developed box potentials \cite{Gaunt2013} make it practical to study homogeneous properties in cold-atom systems. Moreover, a weak background harmonic potential does not change transport properties qualitatively \cite{MCFshort}. Therefore, we will focus on the intrinsic transport phenomena in a homogeneous system with open boundary conditions and employ the microcanonical picture of transport which is ideally suited for closed finite systems \cite{Micro,MCFshort}.

The initial condition is similar to that in Ref.~\cite{MCF_TD}, where the system is separated into two regions (left and right) by a laser sheet, and fermions are only loaded to the left, as illustrated in Fig.~\ref{Fig:Illustration}. We first consider single-species fermionic atoms, which are noninteracting fermions. In Refs.~\cite{MCFshort,MCF_TD,Chern14} it has been shown that the dynamics is quite insensitive to the initial filling as long as the corresponding thermodynamics limit exists, so we consider an initial band insulator on the left half lattice. Adjusting the initial filling should not lead to qualitative differences, except in the flat-band effect discussed later. At time $t=0$ the optical barrier is lifted and the fermions start to propagate to the right. The full quantum dynamics of noninteracting fermions can be monitored by the single particle correlation matrix defined by its elements $C_{ij}(t)=\langle c^{\dagger}_{i}(t)c_{j}(t)\rangle $ \cite{MCFshort} and from this we compute the density on site $i$, $n_i(t)=C_{ii}(t)$.

The equations of motion for $C_{ij}(t)$ can be obtained via
\begin{equation}
i \frac{\partial {\langle c_i^{\dagger} c_j \rangle }}{\partial t} = \langle[  c_i^{\dagger} c_j , H]\rangle = \sum_{\Delta}  (\bar{t}_{i,\Delta}\langle c_{i-\Delta}^{\dagger} c_j \rangle- \bar{t}_{j,\Delta}\langle c^{\dagger}_i c_{j+\Delta}  \rangle).
\end{equation}
Here $\Delta$ denotes the vector to the other side of a link.
The equation of motion was evaluated numerically using the fourth-order Runge-Kutta method \cite{NumRecipes}. The initial correlation matrix was set up with one particle located at each site on the left half of the lattice. We have calibrated the numerical procedure using known exact solutions.

In the thermodynamic limit of infinitely large lattices, the results should agree with band theory \cite{AshcroftBook}. By implementing the lattice version of the Fourier transform, the Hamiltonian becomes
\begin{equation}
H = \sum_k \epsilon_k c_k^\dagger c_k.
\end{equation}
Here $c_{k}$ ($c^{\dagger}_{k}$) is the annihilation (creation) operator in momentum space. The dispersion $\epsilon_{k}$ can be exactly solved for noninteracting fermions. Importantly, the semiclassical group velocity is given by \cite{AshcroftBook}
\begin{equation}
\mathbf{v}_{k}=\nabla_{k}\epsilon_{k}.
\end{equation}
This semiclassical prediction in the thermodynamic limit will be compared to the fully quantum mechanical results of matter-wave propagation.

\section{Result and discussion}\label{sec:result}
We first consider uniform lattices where all hopping coefficients are equal: $\bar{t}_{ij}=\bar{t}$. The time it takes for a particle to tunnel through an isolated double-well potential is $\pi t_0/2$ \cite{Clark04}, and in a 1D multi-site lattice it asymptotically takes $t_0/2$ to tunnel through one link \cite{MCFshort,MCF_TD} because of the modification of the dispersion. Since quantum matter wave is sensitive to how each site is connected to its neighbors, with suitably chosen lattice geometry the propagation speed can be altered.

\begin{figure}
\includegraphics[width = 1.9in]{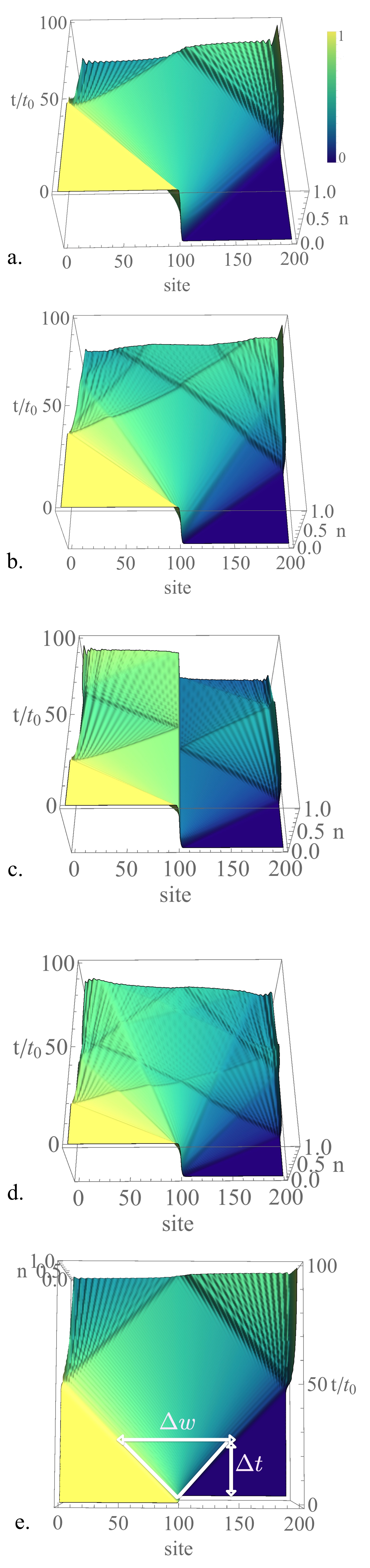}
\caption{(Color online) Evolution of the particle density. Here the brighter regions correspond to higher density. (a) Square lattice. (b) Triangular lattice. (c) Two-leg SNNN lattice. Here $L_x = 200$ and $L_y = 2$. In (c) there is a density discontinuity at $L_x/2$ caused by the flat band. (d) The SNNN lattice with $L_x=200$ and $L_y = 3$ shows no density discontinuity. (e) Using (a) as an example, the spreading of density, $\Delta w/\Delta t$, gives twice the maximal group velocity.}
\label{Fig:Contour}
\end{figure}

Figure~\ref{Fig:Contour} shows the evolution of the $y$-direction averaged density, $\bar{n}(i_x)\equiv (\sum_{i_y=1}^{L_y}n_{i_x,i_y})/L_y$, for the square, triangular, and SNNN lattices with $(L_x,L_y)=(200,2)$ ((a)-(c)) and for the SNNN lattice with $(L_x,L_y)=(200,3)$. At $t=0$ the left half is filled and the right half is empty. The ballistic transport of fermions clearly shows a light-cone structure \cite{Chien14,Ronzheimer13} with the outermost wavefront showing a constant maximal propogating speed. For different geometries the maximal speeds are different, and there are equivalent ways for measuring them. For instance, the expansion rate of the perturbed region of the cloud given by $\Delta w/\Delta t$ will be twice of the maximal propagating speed~\cite{Ronzheimer13} and can be extracted following the illustration of Fig.~\ref{Fig:Contour} (e). Since the ballistic expansion of noninteracting fermions leads to constant maximal propagating speed as shown in Fig.~\ref{Fig:Contour} (a)-(d), measuring the time $t_x$ it takes for the outermost wavefront to reach a point $i_x$ can also be used to evaluate the maximal speed $v_m=(i_x-i_{L/2})/t_x$. In our simulations we impose open boundaries and use the latter method to determine the maximal propagating speed. A density-averaged velocity may be defined following Ref.~\cite{Ronzheimer13}, which works as an indicator of the average velocity of condensed bosons. Since fermions span the momentum space up to the Fermi momentum, the density-averaged velocity is less informative in revealing the spreading velocity and here we focus on the maximal group velocity of fermionic matter waves.

It can be observed that the wavefront speed increases as more diagonal links are added to the lattice. The full quantum dynamics thus claims that altering the lattice geometry may boost matter-wave propagation, which can be very difficult to explain using classical physics. Even more puzzling is that in the case of the SNNN lattice with $L_x=200$ and $L_y=2$, about half of the particles stay on the left half, and this causes a density discontinuity at $L_x/2$, which is visible on Figure~\ref{Fig:Contour}(c). In contrast, for the SNNN lattice with $L_x=200$ and $L_y=3$ no density discontinuity can be observed. This indicates that $L_y=2$ is special for the SNNN lattice.

\begin{figure}
\includegraphics[scale = 0.6]{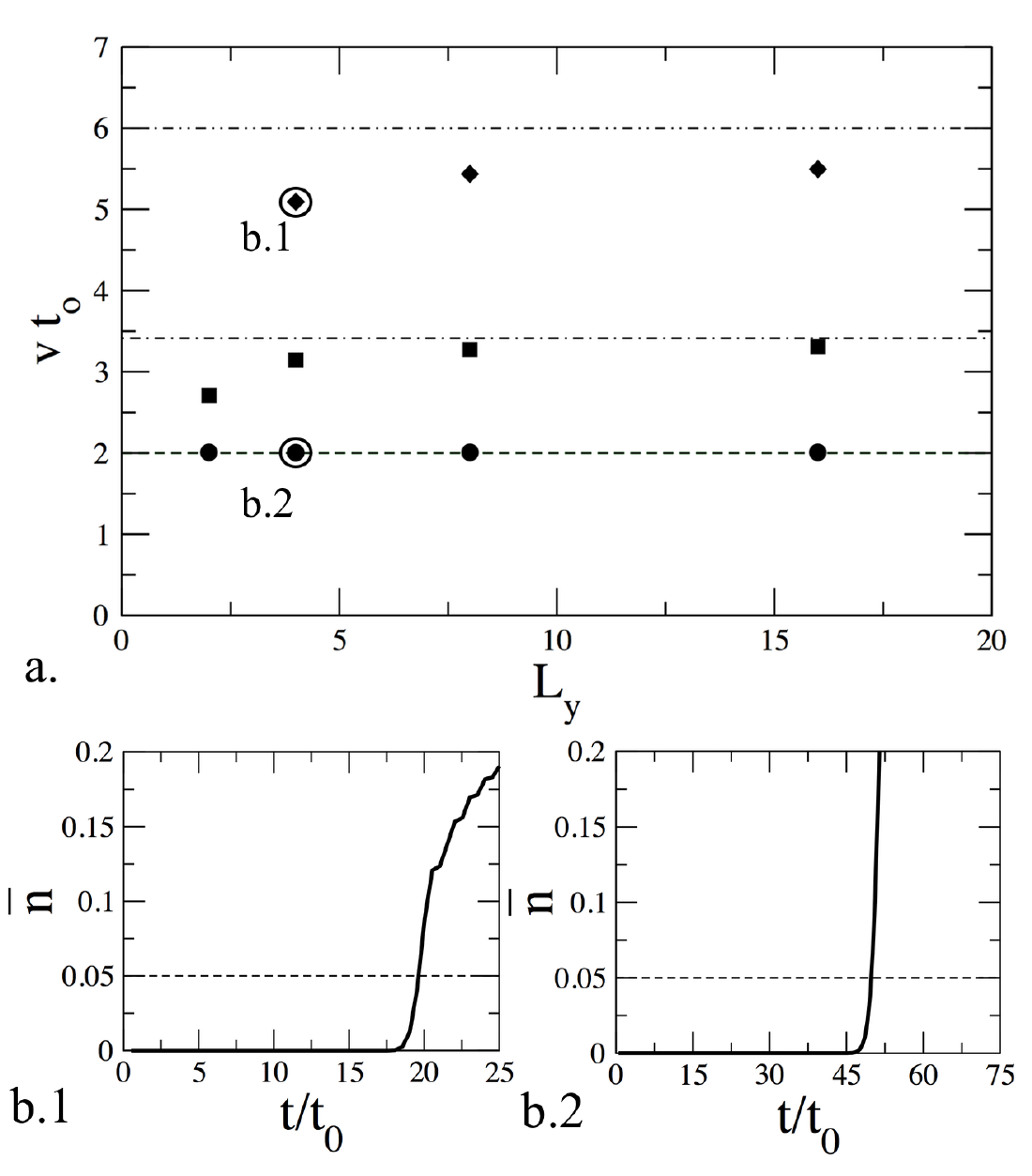}
\caption{(a) Matter-wave propagation velocities for the square (circular symbols), triangular (square symbols), and SNNN (diamond symbols) lattices for $L_x=200$ and selected $L_y$. The horizontal lines show the maximal velocities calculated from band theory. The evolutions of the $y$-direction averaged density $\bar{n}$ for two selected cases indicated as b.1 (for the SNNN lattice) and b.2 (for the square lattice) are shown in panels b.1 and b.2. The wavefront arrival time at $L_x$ (denoted by $t^*$) is determined by the time when $\bar{n}$ rises above a selected threshold. In this work we choose the threshold to be $0.05$ as shown by the dashed lines in panels b.1 and b.2. }
\label{Fig:LyVsSize}
\end{figure}

By analyzing $\bar{n}$ as a function of $t$ as shown in Fig.~\ref{Fig:Contour}, one can estimate the time it takes for the wavefront to first arrive at a point $i_x$. For convenience, we take $i_x=L_x$ and measure $t^*$ as the time when $\bar{n}$ reaches a threshold and extract the velocity $v=(L_x/2)/t^*$ for the three types of lattices shown in Fig.~\ref{Fig:Illustration} as a function of $L_y$ with $L_x=200$ and present the data in Figure~\ref{Fig:LyVsSize} (a). The details of $\bar{n}$ as a function of $t$ for selected cases are shown in Figure~\ref{Fig:LyVsSize} (b1) and (b2). The threshold is set to be $0.05$ and the results do not change qualitatively with this choice. We remark that by measuring the spreading rate of the density profile shown in Fig.~\ref{Fig:Contour} and using the same threshold to determine the boundary of the density spreading, the same maximal propogating speed would be determined. The agreement is guaranteed by the light-cone spreading of noninteracting fermions shown in Fig.~\ref{Fig:Contour}.

\subsection{Band theory}
The geometry-dependent propagation velocity can be accounted for by the standard band theory for fermions \cite{AshcroftBook}. For an infinitely large 2D uniform square lattice following the illustration of Fig.~\ref{Fig:Illustration} (a), we introduce lattice Fourier transform with lattice momentum $k=(k_x,k_y)$ and diagonalize the Hamiltonian \cite{Mahan_CondMatBook}. Then the energy dispersion is 
\begin{equation}
\epsilon_{S}(k) = -2\bar{t}[\cos(k_x) + \cos(k_y)].
\end{equation}
The semiclassical velocity in the $x$-direction is
\begin{equation}
v_x^{S} = 2 \bar{t} \sin(k_x).
\end{equation}
Thus the maximal velocity in the $x$-direction is $2\bar{t}$, which is consistent with the numerical simulations for the square lattice. Since $k_x$ and $k_y$ are decoupled, the result is independent of $k_y$ so the maximal velocity already reaches the band-theory prediction when $L_y=2$, as shown in Fig.~\ref{Fig:LyVsSize} (a).

For the triangular lattice, the energy band is
\begin{equation}
\epsilon_{T}(k) = -2 \bar{t}\left[ \cos(k_x) + 2\cos({k_x \over 2})\cos({\sqrt{3} k_y \over 2}) \right].
\end{equation}
The corresponding velocity is
\begin{equation}
v_x^{T} = 2 \bar{t}\left[\sin(k_x) + \sin({k_x \over 2})\cos({\sqrt{3} k_y \over 2})\right].
\end{equation}
Unlike the square lattice, however, an important feature for this case is that $k_x$ and $k_y$ are mixed in the expressions. The maximal velocity is $2\bar{t}(1 + \sqrt{2}/2)$, which is larger than that in the square lattice. This $k_y$ dependence also suggests that the $L_y$ dependence will be more significant. Indeed, Fig.~\ref{Fig:LyVsSize} (a) shows a stronger $L_y$ dependence of the wavefront propagation velocity for the triangular lattice.

\begin{figure}
\includegraphics[width=3.4in]{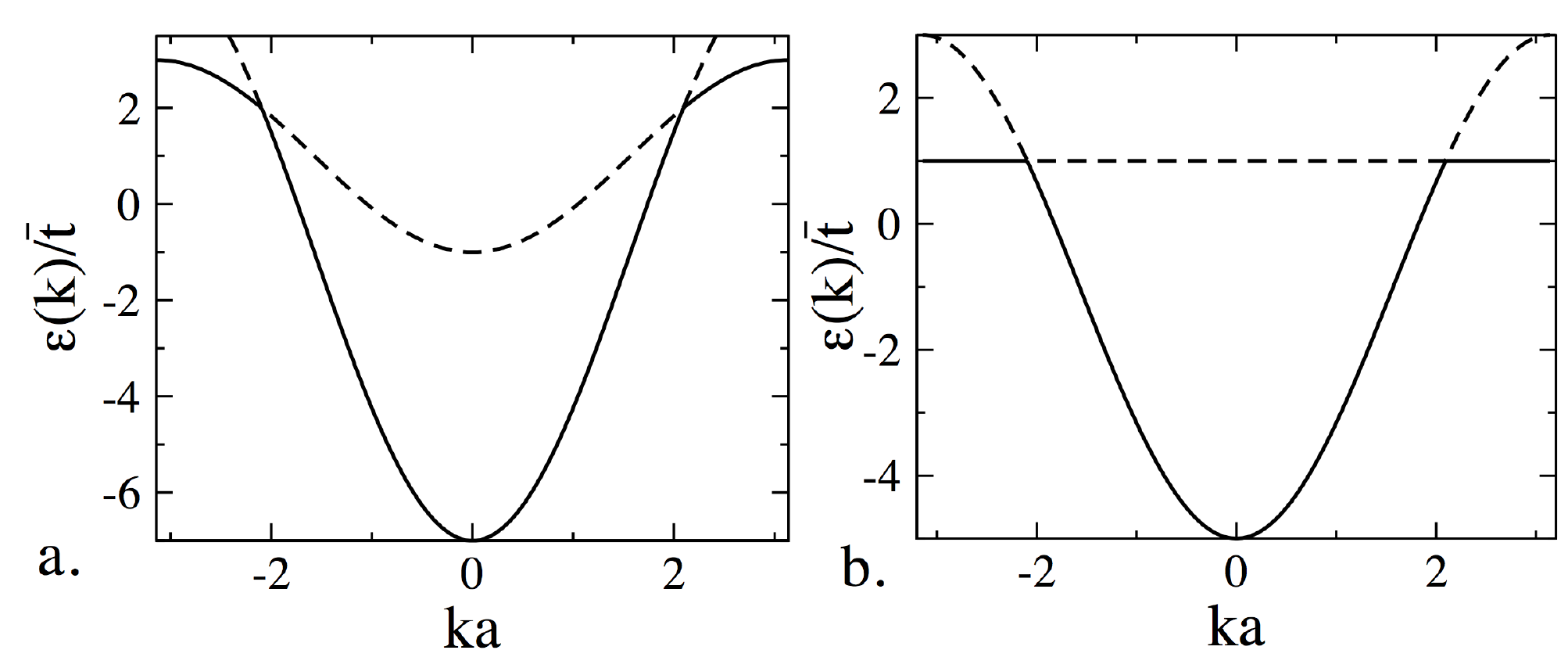}
\caption{Energy bands of (a) 2D SNNN lattice with $k_y = \pi /a$ and (b) two-leg SNNN lattice where a flat band emerges from the overlapping of energy bands when the tunneling coefficients are uniform.}
\label{Fig:EnergyBands}
\end{figure}

For the SNNN lattice, we first consider the infinite 2D lattice and defer the discussion of the special case of $L_y=2$ to later. Since there are two intercalated sublattices (A and B), the energy band will split into two bands while the size of the first Brillouin zone is only half when compared to the previous two cases. The energy dispersion is
\begin{eqnarray}\label{Eq:BCSEk}
\epsilon_{SNNN}(k) &=& -4 \bar{t} \cos(k_x) \pm \bar{t} g(k).
\end{eqnarray}
Here $g(k)=\sqrt{1 + 4\cos(k_x)\cos(k_y) + 4\cos^2(k_x)}$. The group velocity is
\begin{eqnarray}
v_x^{SNNN} &=& 4\bar{t}\sin(k_x)\cos(k_y) \mp \nonumber \\
& &\bar{t}[2\sin(k_x)\cos(k_y)+4\cos(k_x)\sin(k_x)]/g(k). \nonumber \\
& &
\end{eqnarray}
A plot of the two bands at $k_y=\pi/a$ is shown in Figure~\ref{Fig:EnergyBands}. The two bands touch at four points in the first Brillouin zone so the system remains conducting for any filling less than the band insulator. The maximal velocity for the SNNN lattice is $ 6 \bar{t}$. 

As shown in Fig.~\ref{Fig:LyVsSize}, the wavefront velocity of the SNNN lattice approaches the maximal velocity as $L_y$ increases, but the convergence is slower when compared to the previous two cases. Fig.~\ref{Fig:Contour}(d) shows many trailing wakes behind the first wavefront in the SNNN lattice case and they are the byproducts of the more complex dispersion \eqref{Eq:BCSEk}. As a consequence, the averaged density $\bar{n}$ of the SNNN lattice rises less abruptly when compared to the other cases as shown in Fig.~\ref{Fig:LyVsSize} (b.1) and (b.2). By estimating the velocity from the time when $\bar{n}$ rises above the selected threshold $0.05$, an underestimation is in place. This can be improved by choosing a lower threshold of $\bar{n}$, although it would require more precise experiments to locate the lower-density threshold. As illustrated in Fig.~\ref{Fig:Contour}, the same issue may persist if one measures the density-averaged velocity \cite{Ronzheimer13}. The low-density wavefront corresponds to larger distances from the initial boundary, so the product of the density and the distance around the wavefront leads to higher uncertainties in the averaged velocity.

We have two remarks. Our initial condition corresponds to a band insulator on the left half lattice. It has been shown that reducing the initial filling only reduces the magnitude of the current and the dynamics remains similar \cite{MCFshort,Chern14}. One may use an initial configuration with finite total momentum, which may be realizable by filling up fermions in selected momentum states or imposing a momentum transfer. Since the spectral weight of the states are determined by the initial condition, the propagation will be biased towards the selected initial momentum. The maximal propagating speed, however, is determined by the lattice geometry and will not be changed by different initial states. Here we focus on fermionic transport because it has been shown that noninteracting bosons in the ground state will form a macroscopically coherent condensate and the transport is highly oscillatory \cite{MCFshort}. In addition, at low temperature bosons stay in the bottom of energy bands, so they are insensitive to flat-bands which are at higher energies. For the purpose of probing the maximal propagating speed and flat-band physics, fermions are then a better choice.

The honeycomb lattice, when the zig-zag side is placed along the horizontal direction, can be viewed as a deformed square lattice with alternating missing vertical links. In this case, the wavefront propagation velocity from the band theory and simulations are both identical to that of the square lattice when geometrical factors are properly considered. It is known that the low-energy dispersion in the honeycomb lattice around the Dirac points resembles relativistic fermions \cite{Mahan_CondMatBook}. For global transport studied here, the low-energy effects are not observable.

\subsection{Flat band effects}
We now turn to the reason behind the density jump of the two-leg SNNN lattice observed in Fig.~\ref{Fig:Contour}. When $L_y=2$, the Hamiltonian can also be diagonalized and the energy bands are
\begin{equation}
\epsilon(k_x) = -2\bar{t}\cos(k_x) \pm \bar{t}|1 + 2\cos(k_x)|.
\end{equation}
The two bands are shown in Fig.~\ref{Fig:EnergyBands}(b). One important feature is that there is a flat band, i.e., a collection of dispersionless states. The particles on the flat band have localized spatial patterns so they do not participate in transport directly. For the two-leg SNNN lattice, the flat-band state corresponds to the wavefunction $\frac{1}{\sqrt{2}}\left(\begin{array}{r} 1 \\ -1 \end{array}\right)$ localized on a vertical link. The tunneling from the two sites on a verticle link to any site to the left or to the right leads to a total destructive interference, which explains the localization. This also contrasts why $N_y=2$ is special for the SNNN lattice because no such total desctructive interference can be found in other values of $N_y$. In the presence of a chemical potential difference caused by the initial density difference, mobile atoms on the curved part of the dispersions are driven to the initial vacuum region. 

The atoms on the flat-band, in contrast, remain in the initially filled region because their wave functions tend to localize. Since each momentum state can accommodate one fermion, in Fig.~\ref{Fig:EnergyBands}(b) the flat-band contains exactly $L_x$ states. Thus, the flat-band states account for half of the total states (i.e., $L_x$ out of the total $2L_x$ states). Therefore, the static property of the localized flat-band states is the reason for the density jump. By inspecting the particle density profiles at different time slots, we have verified that the density jump does not decrease significantly as the wavefront propagates to the right. We also remark that, because fermions pile up from the lowest energy to the Fermi energy, the lower part of the dispersive band will be populated first. There are $(2/3)L_x$ states on the dispersive band below the energy of the flat-band as shown in Fig.~\ref{Fig:EnergyBands}(b). Thus, to observe the density discontinuity the initial filling on the left half lattice should exceed $1/3$ for the flat-band to be filled. This analysis again shows the power of band theory in explaining transport properties.

Such a flat-band induced dynamical density discontinuity may be generic in cold-atom systems. For instance, the kagome lattice has one flat band in its lowest three bands and quantum dynamics show that a density jump of magnitude $1/3$ emerges as particles flow from an initially filled region into an initially empty region \cite{Chern14}. There is, however, a subtle difference between the flat band of the kagome lattice and that of the two-leg SNNN lattice. The kagome flat-band is on top of the other two dispersive bands while that in the two-leg SNNN intersects the other dispersive band. Nevertheless, their roles in transport are identical. While demonstrating such a dynamical density discontinuity is straightforward in cold-atoms systems, in conventional condensed matter depleting mobile electrons completely can be very challenging and observing this phenomenon can be a daunting task.

\begin{figure}
\includegraphics[width=2.in]{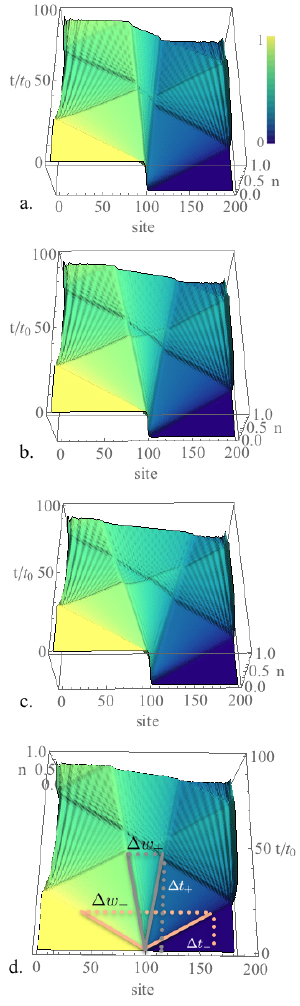}
\caption{Density evolution of two-leg SNNN lattices with $\bar{t}^{\prime}/\bar{t}=0.9$ (a), $0.8$ (b), and $0.7$ (c). Here $L_x=200$. The case with $\bar{t}^{\prime}/\bar{t}=1$ has been shown in Fig.~\ref{Fig:Contour} (c). (d) Using (b) as an example, the maximal rates of density spreading, $\Delta w_{\pm}/\Delta t_{\pm}$, bound the light-cone structures of the two bands.}
\label{Fig:DistortedBand}
\end{figure}

A few remarks on the two-leg SNNN lattice are worth mentioning. First, the existence of the flat band depends crucially on the condition of uniform tunneling coefficients. If the diagonal (A-A and B-B) links have a different tunneling coefficient $\bar{t}^{\prime}$, the bands start to curve and there is no longer a flat band.  We have tested the case for $\bar{t}^{\prime}<\bar{t}$ and found that when $\bar{t}^{\prime}/\bar{t}$ is close to $1$, the flat-band is bent only slightly and a sharp density difference is still observable, but as $\bar{t}^{\prime}/\bar{t}$ decreases, the sharp density difference decays away. The resulting dispersion of the two-leg SNNN lattice
\begin{equation}
 \epsilon(k) = -2\bar{t}\cos(k_x) \pm |\bar{t} + 2\bar{t}^{\prime}\cos(k_x)|
\end{equation}
yields two maximal group velocities for the two bands. As the flat band distorts, the atoms propagate in a light-cone structure bound by $v_{+} = 2(\bar{t}-\bar{t}^{\prime})$, while atoms in the dispersive band reach a maximum velocity of $v_{-} = 2(\bar{t} + \bar{t}^{\prime})$. Figure~\ref{Fig:DistortedBand} summarizes the behavior as $\bar{t}^{\prime}/\bar{t}$ is decreased. We also caution that the density-averaged velocity \cite{Ronzheimer13} may not fully account for the underlying transport when multiple bands are present. For instance, there are two maximal velocities shown in Fig.~\ref{Fig:DistortedBand}, and an averaged velocity will average out this feature.

Secondly, in the presence of magnetic flux penetrating the ladder, the dispersion exhibits additional interesting features and the system is known as the Creutz ladder \cite{MCreutz,CreutzLadder} and transport in the square lattice with effective magnetic flux has been recently studied in cold-atom systems~\cite{Atala14}. Here, we focus only on geometrical effects on transport.

\subsection{Weak interactions}
Transport behavior can be altered by introducing strong interactions as demonstrated in Ref.~\cite{Schneider12}. In the presence of interaction, however, band theory has limited applicability and full numerical simulations are usually required. A lack of efficient numerical methods for studying 2D interacting systems makes a full discussion of interaction effects out of the scope of this work. Here we only demonstrate that in the presence of weak interactions which may be treated in the mean-field approximation, the geometrical effects are still observable. We consider two-component fermions labeled by spins $\sigma=\uparrow,\downarrow$, which may be different species of atoms or the same atoms in two different internal states. For contact interactions common in cold atoms, the system may be described by the Hubbard model \cite{Hofstetter02} $H_{int} = \sum_{\sigma}H_{\sigma} + U \sum_i \hat n_{i\uparrow} \hat n_{i\downarrow}$, where $\hat n_{i\sigma} = c^{\dagger}_{i \sigma}c_{i \sigma}$ and $U$ is the onsite repulsion coupling constant.

The equations of motion for the Hubbard model cannot be solved analytically due to the presence of multi-particle correlations. Here, we follow Ref.~\cite{MCF_TD} and implement the standard Hartree-Fock approximation by decomposing $\langle c^{\dagger}_{i \uparrow}c_{i \uparrow}c^{\dagger}_{i \downarrow}c_{i \downarrow} \rangle$ as $\langle c^{\dagger}_{i \uparrow}c_{i \uparrow}\rangle \langle c^{\dagger}_{i \downarrow}c_{i \downarrow} \rangle$. Then the equations of motion become
$i  \partial_t \langle c_i^{\dagger} c_j \rangle  =  \sum_{\Delta} ( \bar{t}_{j,\Delta}\langle c^{\dagger}_i c_{j+\Delta}  \rangle+ \bar{t}_{i,\Delta} \langle c_{i-\Delta}^{\dagger} c_j \rangle \big) +  U(\langle c^{\dagger}_ic_j \rangle \langle c^{\dagger}_ic_i \rangle - \langle c^{\dagger}_ic_j \rangle \langle c^{\dagger}_jc_j \rangle)$, which can be solved numerically. We impose the condition $\langle c^{\dagger}_{i \uparrow}c_{i \uparrow} \rangle = \langle c^{\dagger}_{i \downarrow}c_{i \downarrow} \rangle$ at any time. Our numerical results for $0<U<\bar{t}$ show no qualitative difference from the corresponding noninteracting fermion cases. On the other hand, the interactions could influence the flat band and there will be leaks from the flat-band states. However, the leaking rate is extremely small in the weakly interacting regime $U/\bar{t}<1$. Within the time scale of our simulations, the density discontinuity remains visible. Therefore, the conclusions from the noninteracting case should hold when the background interactions are weak.

\section{Experimental implications}\label{sec:exp}
\subsection{Possible realization of the SNNN lattice}
Here we discuss possible experimental realizations and applications of this work. The optical square and triangular lattices have been demonstrated \cite{Struck11,Panahi11,Tarruell12,Jo12} and geometrical effects on transport should be observable on those lattices. The SNNN lattice is similar to the checkerboard lattice of Ref.~\cite{Tarruell12} whose diagonal tunneling (A-A or B-B) is expected to be weaker than the nearest-neighbor tunneling (A-B) due to the in-plane geometry. The SNNN lattice with a tunable ratio between those two tunneling coefficients may be created by bilayer square lattices illustrated in Figure~\ref{Fig:Illustration} (e), which is similar to those discussed in Ref.~\cite{LiebLattice}, but the need of an additional phase shift between the two layers will be a challenge. By adjusting either the lattice depth on each sheet or the distance between the two sheets, the relative strength of the two tunneling coefficients can be tuned.

While genuine bilayer optical lattices have not be available, multi-layer optical lattices have been experimentally realized for the honeycomb lattice as an analogue of graphene \cite{MultilayerGraphene}. To realize two-leg ladders of the SNNN lattice for supporting the flat-band, a tight background harmonic trap or laser sheets \cite{Gaunt2013} for confining the atoms in a narrow transverse region may also be needed. As the geometry and tunneling coefficients change, the energy dispersion as well as the group velocity of the fermions change accordingly.

\subsection{Possible application}
\begin{figure}
\includegraphics[width = 3.4in]{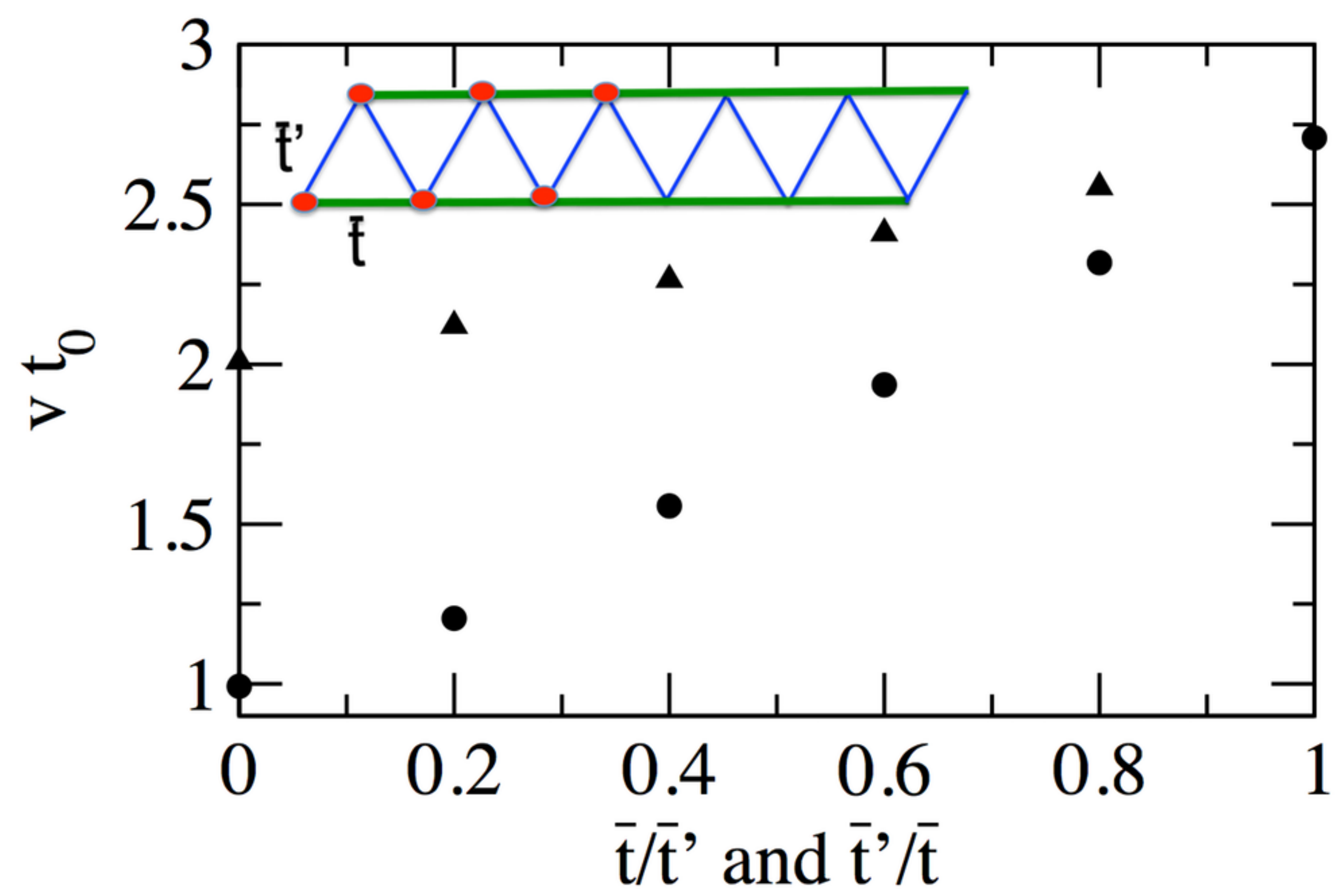}
\caption{(Color online) Matter-wave propagation velocity of a zig-zag lattice (illustrated in the inset) with tunable tunneling coefficients. The zig-zag lattice has two sets of tunneling coefficients: $\bar{t}$ for the horizontal tunneling and $\bar{t}^{\prime}$ for the diagonal tunneling. When $\bar{t}^{\prime}$ is fixed, the velocity increases as the ratio $\bar{t}/\bar{t}^{\prime}$ (circle symbols) increases from $0$ to $1$. When $\bar{t}$ is fixed, the velocity increases as $\bar{t}^{\prime}/\bar{t}$ (triangular symbols) increases from $0$ to $1$.  A maximal group velocity emerges when $\bar{t}^{\prime}/\bar{t}=1$.}
\label{Fig:ZZL}
\end{figure}
We end our discussion of geometrical effects on matter-wave propagation by presenting a possible application in the zig-zag lattice optical lattice \cite{ZigZag,Jo15}, which is basically a two-leg ladder of the triangular lattice as shown in the inset of Figure~\ref{Fig:ZZL}. The hopping coefficient along the horizontal direction is $\bar{t}$ and that along the diagonal is $\bar{t}^{\prime}$, and the ratio between them are assumed to be continuously tunable. The initial condition is identical to the previous cases with the left half of the lattice being a band insulator and the right half being empty, as illustrated in Fig.~\ref{Fig:ZZL}.

When $\bar{t}^{\prime}=0$ and $\bar{t}$ gives the time scale $t_0$, the two legs decouple and it takes $L_xt_0/4$ for the matter wave to reach the far right boundary. In the opposite limit when $\bar{t}=0$ and $\bar{t}^{\prime}$ gives the same time scale $t_0$, it takes $L_xt_0/2$ because the particle has to traverse twice the distance to reach the far right boundary. The velocities are thus $2a/t_0$ and $a/t_0$ for the two limits. Figure~\ref{Fig:ZZL} shows the velocity for selected values of $\bar{t}/\bar{t}^{\prime}$ and $\bar{t}^{\prime}/\bar{t}$. We chose the quantity in the denominator to give the same tunneling time scale $t_0$ to guarantee a fair comparison. Importantly, a maximal velocity appears when $\bar{t}=\bar{t}^{\prime}$ when each site has the most available number of neighbors to tunnel into.

By tuning the ratio of the tunneling coefficients, a zig-zag optical lattice may serve as an atomic ``bicycle gearing'' for changing the speed of atoms passing through it. One possible scenario is to control the nearest-neighbor tunneling $\bar{t}$ while keep the next-nearest-neighbor tunneling $\bar{t}^{\prime}$ fixed. When $\bar{t}=0$, the velocity is slowed down to $a/t_0$. To accelerate the matter wave, one can tune to $\bar{t}=\bar{t}^{\prime}$ and the velocity will be increased by $270\%$, according to Fig.~\ref{Fig:ZZL}. The zig-zag lattice tunes matter-wave propogating speed by the relative hopping strength of the two types of links (horizontal and diagonal) while the total current through the lattice is roughly maintained in a finite range because the total bandwidth does not change drastically. One may consider an alternative way of simply reducing all the hopping coefficients of a lattice to decrease the overall propagating speed, but this will reduce the total current due to the reduced bandwidth as well. An experimental set up proposed for the zig zag lattice by superpositioning a triangular lattice and a strong optical superlattice \cite{BosonsTriangle, OpTriangleLattice,Jo15} could make such a gearing device feasible.

Similar applications may be relevant to quantum information transfer by controlling the group velocity of information carriers played by  massive quantum particles. There have been experiments demonstrating information exchanges between photons and cold-atom matter-waves \cite{Hau2007_MWInfo}, where the propagtion speed of the matter wave affects the overall transfer of the information. Our investigations into geometrical control of matter-wave propagation could provide addtional manipulations of quantum systems.

\section{Conclusions}\label{sec:conclusion}
We have explored geometrical effects on quantum transport of cold atoms. With different energy dispersions from various lattice geometries, significantly increases of matter-wave propagation should be observable. With the assistance of band theory, searching and implementing lattice geometries for controlling transport can be performed efficiently. Moreover, insulating phases sustaining a density difference is a generic feature for lattice geometries supporting a flat band, and the different roles of mobile and localized atoms in quantum transport have been elucidated. Similar studies can be performed in the recently realized optical Lieb lattice~\cite{Taie15} and the proposed optical saw-tooth lattice~\cite{Jo15}, which also support flat-bands.

Optimal designs of lattice geometries for controlling and regulating quantum transport of cold atoms could contribute useful elements to the thriving field of atomtronics \cite{Atomtronics07, Atomtronics09} for simulating or complementing electronics. The feasibility of tuning  matter-wave propagation using geometrical effects may also find applications in quantum quench dynamics \cite{CAdyn2,Dziarmaga10}, where the relaxation rate after a quench may be controlled by geometrical effects.

\textit{Acknowledgment - }
We thank Michael Zwolak, Kevin Mitchell, Fei Zhou, and Dan Stamper-Kurn for useful discussions, and Michael Colvin and Chen-Yen Lai for assisting the numerical calculations. G. W. C acknowledges the support of the U. S. DOE through the LANL/LDRD Program. MD acknowledges support from the DOE Grant No. DE-FG02-05ER46204.

\bibliographystyle{apsrev4-1}
%

\end{document}